# An early warning tool for predicting mortality risk of COVID-19 patients using machine learning

**Muhammad E. H. Chowdhury[1]*, Tawsifur Rahman[2], Amith Khandakar[1], Somaya Al-Madeed[3], Susu M. Zughaier[4], Suhail A. R. Doi[5], Hanadi Hassen[3], Mohammad T. Islam[6]**

[1]Department of Electrical Engineering, Qatar University, Doha-2713, Qatar

[2]Department of Biomedical Physics & Technology, University of Dhaka, Dhaka-1000, Bangladesh

[3]Department of Computer Science and Engineering, Qatar University, Doha-2713, Qatar

[4]Department of Basic Medical Sciences, College of Medicine, QU Health, Qatar University, Doha-2713, Qatar

[5]Department of Population Medicine, College of Medicine, QU Health, Qatar University, Doha-2713, Qatar

[6] Dept. of Electrical, Electronics and Systems Engineering, Universiti Kebangsaan Malaysia, Bangi, Selangor 43600, Malaysia

*Correspondence: Muhammad E. H. Chowdhury; mchowdhury@qu.edu.qa, Tel.: +974-31010775

## Abstract

### *Background*

COVID-19 pandemic has created an extreme pressure on the global healthcare services. Fast, reliable and early clinical assessment of the severity of the disease can help in allocating and prioritizing resources to reduce mortality.

### *Methods*

In order to study the important blood biomarkers for predicting disease mortality, a retrospective study was conducted on 375 COVID-19 positive patients admitted to Tongji Hospital (China) from January 10 to February 18, 2020. Demographic and clinical characteristics, and patient outcomes were investigated using machine learning tools to identify key biomarkers to predict the mortality of individual patient. A nomogram was developed for predicting the mortality risk among COVID-19 patients.

### *Results*

Lactate dehydrogenase, neutrophils (%), lymphocyte (%), high sensitive C-reactive protein, and age - acquired at hospital admission were identified as key predictors of death by multi-tree XGBoost model. The area under curve (AUC) of the nomogram for the derivation and validation cohort were 0.961 and 0.991, respectively. An integrated score (LNLCA) was calculated with the corresponding death probability. COVID-19 patients were divided into three subgroups: low-, moderate- and high-risk groups using LNLCA cut-off values of 10.4 and 12.65 with the death probability less than 5%, 5% to 50%, and above 50%, respectively.

### *Conclusions*

The prognostic model, nomogram and LNLCA score can help in early detection of high mortality risk of COVID-19 patients, which will help doctors to improve the management of patient stratification.

## INTRODUCTION

The novel coronavirus disease (COVID-19) spread rapidly throughout the world from Wuhan (Hubei, China) since December 2019 [1-5]. Since the outbreak, the number of reported cases has surpassed 12 million with more than 550 thousand deaths worldwide as of 12 July 2020 [6]. The COVID-19 disease is caused by the severe acute respiratory syndrome coronavirus 2 (SARS-CoV-2), which is a member of the coronavirus family. On 11 March 2020, COVID-19 was declared as a pandemic by the World Health Organization (WHO) [7]. Due to the pandemic, hospital capacity is being exceeded in many places and face issues in terms of limited medical staff, personal protective equipment, life-support equipment and others [8, 9]. Symptoms of COVID-19 are non-specific, and infected individuals may develop fever (83-99%), cough (59-82%), loss of appetite (40-84%), fatigue (44-70%), shortness of breath (31-40%), coughing up sputum (28-33%) or muscle aches (11-35%) [10]. The disease can further progress into a severe pneumonia, acute respiratory distress syndrome (ARDS), myocardial injury, sepsis, septic shock, and even death [11]. Though most COVID-19 patients have a mild illness, there are some patients who show rapid deterioration (particularly within 7–14 days) from the onset of symptoms into severe COVID-19 with or without ARDS [12, 13]. Current epidemiological data suggest that the mortality rate of patients with severe COVID-19 is higher than that of patients with non-severe COVID-19 [14, 15]. It has been reported that 26.1-32.0% of patient infected with COVID-19 are prone to progressing critical illness [16]. Recent studies have confirmed a high fatality rate of 61.5% for patients in critical cases, which increases with age and other medical comorbidities [16].

A large cohort study from 2449 patients has reported that during this pandemic healthcare system can be overwhelmed by hospitalization (20-31%) and intensive care unit (ICU) admission rates (4.9-11.5%) [17]. This can be avoided by prioritizing hospital treatment for patients at high risk of deterioration and death, and treating low-risk patients in ambulatory environments, or by home-based self-quarantine. An effective tool is required to predict the disease trajectory to allocate resources efficiently and also improve the patient's condition. Understanding the great potential of this approach, it is important to identify key patient variables that can help to predict the course of the disease at diagnosis. In other words, early identification of patients at high risk for progression to severe COVID-19 will help in efficient utilization of healthcare resources via patient prioritization to reduce the mortality rate.

Several researches indicate that biomarkers can help to classify COVID-19 patients with elevated risk of serious disease and mortality by providing crucial information regarding the patients' health status. Al Youha et al. [18] proposed a prognostic model called the Kuwait Progression Indicator (KPI) Score for predicting progression of severity in COVID-19. The KPI model was based on quantifiable laboratory readings unlike other self-reported symptoms and other subjective parameters based scoring systems. The KPI score categorizes patients to low risk if the score goes below -7 and high risk if the score goes above 16, however, the progression risk in the intermediate group (for patients scores within -6 to 15) deemed by the authors as uncertain. This intermediate category however exists with many prognostic systems. Weng et al. [19] reported an early prediction score called ANDC to predict mortality risk for COVID-19 patients using 301 adult patients' data. LASSO regression has identified age, neutrophil-to-lymphocyte ratio (NLR), D-dimer, and C-reactive protein recorded during admission as mortality predictors for COVID-19 patients [19]. They have developed a nomogram demonstrating good performance and also derived an integrated score, ANDC, with its corresponding death probability. They have also developed cutoff ANDC values to classify COVID-19 patients into three groups: Low, Moderate and High-risk groups. The death probability were 5%, 5% to 50% and more than 50% in the low-, moderate- and high-risk group, respectively. Using a cohort of 444 patients, Xie et al. [20] proposed a prognostic model using lactate dehydrogenase, lymphocyte count, age, and SpO2 as key-predictors of COVID-19 related death. The model showed good discrimination for internal and external validation with C-statistics of 0.89 and 0.98 respectively. (c=0·98) validation. Even though the model shows promising performance for internal calibration, however, external validation showed over and under-prediction for low-risk and high-risk patients respectively.

Yan et al. [21] reported a machine learning approach to select three biomarkers (lactic dehydrogenase (LDH), lymphocyte and high-sensitivity C-reactive protein (hs-CRP)) and using them to predict individual patients mortality, 10 days ahead with more than 90 percent accuracy. In particular, high levels of LDH alone have been found to play a crucial role in identifying the vast majority of cases, which require immediate medical attention. However, there is no scoring system reported in this work, which can help the clinicians to identify the patients under risk quantitatively.

Another clinical study on 82 COVID-19 patients showed that respiratory, cardiac, hemorrhage, hepatic, and renal injury had caused the death of 100%, 89%, 80.5%, 78.0%, and 31.7% patients respectively. Most of the patients had increased CRP (100%) and D-dimer (97.1%) [22]. The value of D-dimer as a prognostic factor was also shown to significantly increase odds of death if the amount is greater than 1 $\mu g\ mL^{-1}$ upon admission [23, 24].

Although several predictive prognostic models are proposed for the early detection of individuals at high risk of COVID-19 mortality, a major gap remains in the design of state-of-the-art interpretable machine learning based algorithms and high performance quantitative scoring system to classify the most selective predictive biomarkers of patient death. Identifying and prioritizing those at severe risks is important for both resource planning and treatment therapy.

Moreover, the high risk patients should be possible to continuously monitored using a reliable scoring tool during their hospital stay-time. Likewise, reducing patient admission with very low risk of complications that can be handled safely by self-quarantine will help to minimize the pressure on healthcare facilities.

Therefore, using state-of-the-art machine learning algorithm, an early prediction scoring system was developed and also implemented to classify the most discriminatory biomarkers of patient mortality. The problem was initially introduced as a classification problem for determining the most appropriate biomarkers at the end of the test period with the aid of corresponding survival or death outcomes. The top ranked features with the best classification performance were used to develop a multivariable logistic regression-based nomogram and validated for the prognosis of death and survival. The findings obtained through this study provides a simple, easy-to-use and reliable algorithm for the prognosis of high-risk individuals and possess potential for clinical application.

## METHODLOGY

### A. HUMAN SUBJECTS AND STUDY DESIGN

Blood samples collected between 10 January and 18 February, 2020 from 375 patients in Wuhan, China were retrospectively analyzed to identify reliable and relevant markers of mortality risk. Medical records were collected using standard case report forms, which included information on epidemiological, demographic, clinical, laboratory and mortality outcomes. Yan et al. [21] has published the dataset along with the article and the original study was approved by the Tongji Hospital Ethics Committee. Patients' exclusion criteria for the study were: Age (<18 years), pregnant, breast-feeding and missing data (>20%). Out of 375 patients, 187 (49.9%) had fever while cough, fatigue, dyspnea, chest distress and muscular soreness were present in 52 (13.9%), 14 (3.7%), 8 (2.1%), 7 (1.9%) and 2 (0.5%) patients respectively.

### B. STATISTICAL ANALYSIS

Stata/MP 13.0 software was used for conducting the statistical analysis. Gender variation was described using number and percentage. Continuous variables, age and other biomarkers were reported with the number of missing data, median, mean, and quartiles (Q1, Q3) for each biomarkers in death, and survival groups. Wilcoxon tests were conducted for all continuous variable while the chi-squared test for univariate analysis such as gender. Statistically significant difference was defined as a P-value <0.05. There were 76 biomarkers present in the original dataset however 14 biomarkers using two-different algorithms were identified as promising and are summarized in Table 1. These 14 biomarkers selected included lactate dehydrogenase (LDH), neutrophils (%), lymphocyte (%), high sensitivity C-reactive protein (hs CRP), serum sodium, eosinophil (%), serum chloride, monocyte (%), international normalized ratio (INR), activated partial thromboplastin time (APTT), high sensitivity cardiac troponin

I, brain natriuretic peptide precursor (NT-proBNP), albumin,

and mean corpuscular hemoglobin concentration (MCHC).

**Table 1:** Statistical Analysis of the Characteristic of the subjects' data

| Item | Survived | Death | Total | Method | Statistic | P value |
|---|---|---|---|---|---|---|
| **Gender** | | | | Chi-square test | $X^2$=21.70 | <0.00001 |
| • **Male (%)** | 98(49%) | 126(72%) | 224(60%) | | | |
| • **Female (%)** | 103(51%) | 48(28%) | 151(40%) | | | |
| **Age** | | | | Rank-sum test | Z=-11 | <0.0001 |
| • **N(missing)** | 201(0) | 174(0) | 375(0) | | | |
| • **Mean ± SD** | 50.2±15 | 68.8±11.8 | 58.8±16.5 | | | |
| • **Median** | 51 | 69 | 62 | | | |
| • **Q1, Q3** | 37, 62 | 62.2, 77 | 46, 70 | | | |
| • **Min, Max** | 18, 88 | 19, 95 | 18, 95 | | | |
| **Lactate dehydrogenase** | | | | Rank-sum test | Z=-13.18 | <0.0001 |
| • **N(missing)** | | | | | | |
| • **Mean ± SD** | 193(8) | 163(11) | 356(19) | | | |
| • **Median** | 271±102 | 642±341 | 441±305 | | | |
| • **Q1, Q3** | 250 | 567 | 336 | | | |
| • **Min, Max** | 203, 312 | 428, 762 | 239, 564 | | | |
| | 119, 799 | 188,1867 | 119, 1867 | | | |
| **Neutrophils (%)** | | | | Rank-sum test | Z=-12.88 | <0.0001 |
| • **N(missing)** | 194(7) | 162(12) | 356(19) | | | |
| • **Mean ± SD** | 65.7±13.8 | 87±9.86 | 75.4±16.1 | | | |
| • **Median** | 66.2 | 89.5 | 77.5 | | | |
| • **Q1, Q3** | 56.5, 75.4 | 83.2, 93.7 | 64.3, 89.2 | | | |
| • **Min, Max** | 1.7, 95.1 | 18.2, 98.7 | 1.7, 98.7 | | | |
| **Lymphocyte (%)** | | | | Rank-sum test | Z=11.97 | <0.0001 |
| • **N(missing)** | 194(7) | 162(12) | 356(19) | | | |
| • **Mean ± SD** | 24.8±11.4 | 7.6±6.22 | 17±12.7 | | | |
| • **Median** | 23.8 | 5.8 | 14.4 | | | |
| • **Q1, Q3** | 16.6, 33.5 | 3.3, 10.1 | 6.1, 25.2 | | | |
| • **Min, Max** | 4.1, 60 | 0, 44.3 | 0, 60 | | | |
| **High sensitivity C-reactive protein** | | | | Rank-sum test | Z=-11.93 | <0.0001 |
| • **N(missing)** | 194(7) | 159(15) | 353(22) | | | |
| • **Mean ± SD** | 36±44 | 127±75.5 | 77±75.4 | | | |
| • **Median** | 19 | 114 | 53 | | | |
| • **Q1, Q3** | 4, 50 | 62, 179 | 12, 118 | | | |
| • **Min, Max** | 0, 237 | 4, 320 | 0, 320 | | | |
| **Serum sodium** | | | | Rank-sum test | Z=-1.57 | 0.12 |
| • **N(missing)** | 193(8) | 161(13) | 354(21) | | | |
| • **Mean ± SD** | 138.9±3.38 | 139.9±8.37 | 139.3±6.18 | | | |
| • **Median** | 139.2 | 138.9 | 139 | | | |
| • **Q1, Q3** | 136.6, 141 | 135.8, 143 | 136.3, 142 | | | |
| | 125, 146.4 | 115.4, 179 | 115.4, 179 | | | |
| • **Min, Max** | | | | | | |
| **Eosinophil (%)** | | | | Rank-sum test | Z=6.63 | <0.0001 |
| • **N(missing)** | 194(7) | 162(12) | 356(19) | | | |
| • **Mean ± SD** | 0.7±.941 | 0.11±0.38 | 0.44±.79 | | | |
| • **Median** | 0.3 | 0.00 | 0.00 | | | |
| • **Q1, Q3** | 0, 1.1 | 0.0, 0.0 | 0.00,0.53 | | | |
| • **Min, Max** | 0, 6.40 | 0, 3.70 | 0.00, 6.40 | | | |
| **Serum chloride** | | | | Rank-sum test | Z=-0.65 | 0.52 |
| • **N(missing)** | 193(8) | 161(13) | 354(21) | | | |
| • **Mean ± SD** | 100.8±3.8 | 101.5±8.56 | 101.1±6.42 | | | |
| • **Median** | 101.3 | 100.6 | 101.1 | | | |
| • **Q1, Q3** | 98.8, 103.3 | 97.1, 105.5 | 97.9, 103.9 | | | |
| • **Min, Max** | 85.6, 109.1 | 71.5, 140 | 71.5, 140 | | | |
| **Monocyte (%)** | | | | Rank-sum test | Z=8.42 | <0.0001 |
| • **N(missing)** | 194(7) | 152(12) | 356(19) | | | |
| • **Mean ± SD** | 8.4±3.15 | 5.1±4.31 | 6.9±4.08 | | | |
| • **Median** | 8.2 | 4 | 6.8 | | | |
| • **Q1, Q3** | 6.6, 10.1 | 2.4, 6.3 | 3.8, 9.2 | | | |
| • **Min, Max** | 0.7, 15.8 | 0.3, 35.2 | 0.3, 35.2 | | | |
| **International standard ratio** | | | | Rank-sum test | Z=-9.4 | <0.0001 |
| • **N(missing)** | 189(12) | 163(11) | 352(23) | | | |
| • **Mean ± SD** | 1.055±.086 | 1.37(1.01) | 1.2±.709 | | | |
| • **Median** | 1.040 | 1.22 | 1.1 | | | |
| • **Q1, Q3** | 1, 1.1 | 1.1, 1.37 | 1, 1.2 | | | |
| • **Min, Max** | 0.84, 1.33 | 0.88, 13.48 | 0.8, 13.5 | | | |
| **Activation of partial thromboplastin time** | | | | Rank-sum test | Z=-1.2 | 0.23 |
| • **N(missing)** | 165(36) | 133(41) | 298(77) | | | |
| • **Mean ± SD** | 40.1±5.7 | 41.9±11.4 | 41±8.7 | | | |
| • **Median** | 39.9 | 39.4 | 40 | | | |
| • **Q1, Q3** | 35.9, 43.5 | 35, 45.4 | 36, 44 | | | |
| • **Min, Max** | 22, 56.9 | 25.3,137 | 22, 137 | | | |
| **Hypersensitive cardiac troponin I** | | | | Rank-sum test | Z=-5.82 | <0.0001 |
| • **N(missing)** | 141(60) | 146(28) | 287(88) | | | |
| • **Mean ± SD** | 12±53.3 | 1391±5748 | 714±414 | | | |
| • **Median** | 3 | 41 | 11 | | | |
| • **Q1, Q3** | 2, 7 | 15, 271 | 3, 50 | | | |
| • **Min, Max** | 2, 617 | 2, 50000 | 2, 50000 | | | |
| **Brain natriuretic peptide precursor (NT-proBNP)** | | | 267(108) | Rank-sum test | Z=-3.87 | <0.0001 |
| • **N(missing)** | 128(73) | 139(35) | 1959±6308 | | | |
| • **Mean ± SD** | 1039±6620 | 2806±5906 | 271 | | | |
| • **Median** | 65 | 827 | 68, 935 | | | |
| • **Q1, Q3** | 23, 178 | 362, 2402 | 5,70000 | | | |
| • **Min, Max** | 5, 70000 | 24, 45850 | | | | |
| **Albumin** | | | | Rank-sum test | Z=10.64 | <0.0001 |
| • **N(missing)** | 193(8) | 163(11) | 356(19) | | | |
| • **Mean ± SD** | 37.1±4.53 | 30.3±4.22 | 34±5.57 | | | |
| • **Median** | 37.4 | 30.1 | 34.2 | | | |
| • **Q1, Q3** | 34.2, 40.2 | 27.6, 33 | 29.9, 38.3 | | | |
| • **Min, Max** | 22.6, 48.6 | 18.5, 40.9 | 18.5, 48.6 | | | |
| **Mean corpuscular hemoglobin** | | | | Rank-sum test | Z=-2.27 | 0.023 |
| **concentration** | | 162(12) | 356(19) | | | |
| • **N(missing)** | 194(7) | 346±18.7 | 345±16.3 | | | |
| • **Mean ± SD** | 343±13.9 | 346 | 345 | | | |
| • **Median** | 344 | 337,354 | 336, 352 | | | |
| • **Q1, Q3** | 335, 351 | 299,488 | 299, 488 | | | |
| • **Min, Max** | 306,416 | | | | | |
| **Outcome (%)** | 201(54%) | 174(46%) | 375 | | | |

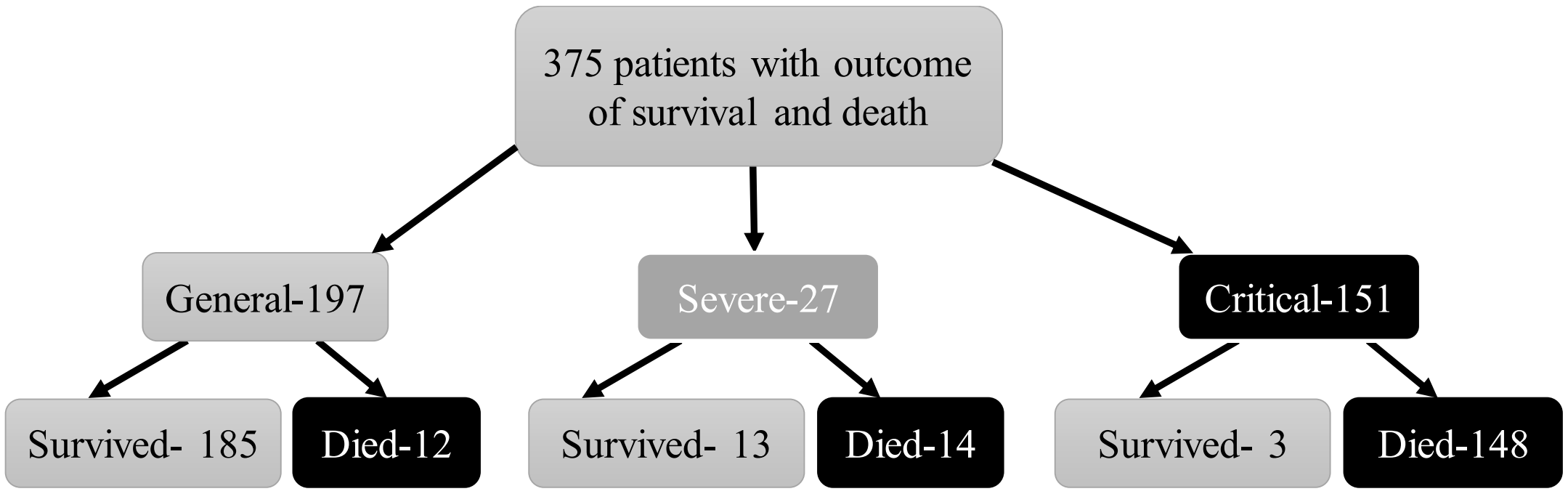

**Figure 1:** Patients' outcome tree with the initial condition of the patients in admission.

### C. *FEATURE RANKING*

Even though multiple blood sample data of the patients were available, only the data from the first sample were used as inputs for model training and validation to identify the key predictors of the disease severity. The model also helps in distinguishing patients that require immediate medical assistance. Research using clinically captured data often suffers from missing data challenge leading to either bias introduction or negative impact on analytical outcomes. Simple approach to handle this challenge is deleting the respective rows of data from further analysis. This simple approach is not very useful as it leads to loss of valuable information that would have been beneficial in the analysis and also can lead to biased estimates [25].

Multiple imputation using chained equations (MICE) data imputation technique is the most popular technique for clinical data imputation. MICE technique uses multiple regression models to predict the missing data depending on other variables in the dataset. In this technique, each of the missing variables is modelled depending on the datatype. Binary variables are predicted using logistic regression while continuous variables were predicted using predictive mean matching [25].

In this study two different data imputation techniques are compared. In the first technique [21], missing data were padded by '−1' and normalized by 'Z-score' whereas in the second technique, missing data were imputed using MICE technique and normalized using 'Z-score'.

Each of the 76 parameters were assessed to take decisions and identify the top-14 biomarkers in addition to age and gender, to obtain the top-ranked biomarkers as mortality predictors. Two different sets (Top 10 features) were identified using Multi-Tree Extreme Gradient Boost (XGBoost) technique [26], according to their importance from the imputed and normalized data using two imputation techniques mentioned earlier. The importance of each individual feature in XGBoost is from its accumulated use in each decision step in trees. The approach is extremely useful when dealing with clinical parameters [21]. Initially, default settings of XGBoost was used, i.e. maximum depth = 4, learning rate = 0.2, tree estimators = 150, regularization parameter α = 1 and 'subsample' and 'colsample bytree' both set to 0.9 to avoid overfitting for cases with many features and limited sample size [21].

### D. ***DEVELOPMENT AND VALIDATION OF THE LOGISTIC REGRESSION MODEL IN CLASSIFYING THE OUTCOME***

This study uses a supervised logistic regression classifier [27] as the predictor model. Logistic regression is a common model used in medical statistics and is a statistical learning technique categorized in supervised' machine learning (ML) methods dedicated to classification tasks [28]. The logistic function is a sigmoid function and shrinks real value continuous inputs into a probability. They also make the independent values more resistant to deviations from normality and thus more consistent coefficients [28] .

ROC curves using testing data were constructed to calculate the area under curve (AUC) for single predictors separately and also combination of them. In order to evaluate the performance of different top ranked features in classifying death and survival cases. The logistic regression classifier was evaluated for different combinations of features as input to the model. The trained algorithms were validated using 5-fold cross-validation (80% data were used for training and validation while remaining 20% data were used for testing and this is repeated 5-times). The performance of different models were evaluated using several performances metrics including sensitivity, specificity, positive likelihood ratio (PLR) and negative likelihood ratio (NLR) using testing dataset. Per-class values were computed over the overall confusion matrix that accumulates all test (unseen) fold results of the 5-fold cross-validation.

$$Sensitivity_{class_i} = \frac{TP_{class_i}}{TP_{class_i} + FN_{class_i}} \quad (1)$$

$$Specificity_{class_i} = \frac{TN_{class_i}}{TN_{class_i} + FP_{class_i}} \quad (2)$$

$$PLR_{class_i} = \frac{Sensitivity_{class_i}}{1 - Specificity_{class_i}} \quad (3)$$

$$NLR_{class_i} = \frac{1 - Sensitivity_{class_i}}{Specificity_{class_i}} \quad (4)$$

where $class_i = Suvival\ and\ Death.$

### E. DEVELOPMENT AND VALIDATION OF LOGISTIC REGRESSION-BASED NOMOGRAM IN THE OUTCOME PREDICTION

A diagnosis nomogram was constructed by Alexander Zlotnik's Nomolog [29], based on multivariate logistic regression analysis, using Stata/MP software version 13.0. Logistic (logit) regression estimates the parameter in the form of a binary regression. Logistic regression works with probability, odds and regression. In the binary logistic model, there is an outcome/indicator variable which has two possible values. The outcome variable is a dependent variable which is typically labeled as '0' and '1' and '0' represent survival and '1' represents death in this case. The odds are the ratio of the probability of an event happening to the probability of not happening. Although the probability can vary between 0 and 1, the odds can vary between 0 and ∞. In logistic regression, the logarithm of odds is a linear combination of one or more independent variables ("predictors") which can be a binary variable (e.g., gender) and continuous variable (e.g., age). The log-odds can be termed as linear prediction (LP) and can be related to the probability of a particular outcome. The following equations were used to create relationship between death probability and the key-predictors using logistic regression:

$$odds = \frac{P}{1-P} \quad (5)$$

$$LP = \ln(odds) = \ln\left(\frac{P}{1-P}\right)$$
$$= b_0 + b_1x_1 + b_2x_2 + \cdots + b_nx_n \quad (6)$$

$$\frac{P}{1-P} = e^{b_0+b_1x_1+b_2x_2+\cdots+b_nx_n} = e^{LP} \quad (7)$$

$$P = \frac{e^{LP}}{1+e^{LP}} = \frac{1}{1+e^{-LP}} \quad (8)$$

The top-ranked features (independent variables) showing best AUC was used for creating the logistic regression based nomogram. The entire dataset was divided into training (70%) and validation (30%) sets. Calibration curves for internal (with development set) and external (with validation set) validation were plotted to compare predicted and actual death probability of patients with COVID-19. Decision curve analysis (DCA) was carried out to identify the threshold values in which nomograms were clinically useful, using Stata software.

### F. DEVELOPMENT AND VALIDATION OF EARLY WARNING SCORE

The parameters were drawn as a numerated horizontal axis scale and the values for the patient are put on the numerated scale. A vertical line was drawn down from the different parameter numerated arranged scales downward to a score axis. All five scores on the score axis were added to make a total score and this was linked to a death probability. It can be noted that according to the nomogram, higher score corresponds to a higher death probability. The model was designed using the initial blood sample of the patients. However, it can be applied to the biomarkers collected in later during the hospital stay period of the patients to predict death probability longitudinally using the LNLCA score.

## RESULTS

### A. *DEMOGRAPHIC CHARACTERISTICS, CLINICAL CHARACTERISTICS, AND CLINICAL OUTCOMES*

Of the 375 patients, 174 (46.4%) died, while 201 (53.6%) patients recovered from COVID-19 and were discharged from hospital. Figure 1 summarizes the outcome of patients based on their initial conditions: general (197), severe (27) and critical (151). The minimal, maximal and median follow-up times (from hospital admission to death or discharge) for all 375 patients are 0 days, 35 days and 12 days, respectively.

Table 1 summarizes the demographic characteristics, clinical characteristics, and clinical outcomes of the subjects in the death and survival groups. There were 142 (37.9%) patients, who were Wuhan residents, 2 (0.5%) had contact with confirmed or suspected patients, 24 (6.4%) were from familial cluster, 7 (1.9%) were health workers, 2 (0.5%) had contact with Huanan Seafood Market and 198 (52.5%) had no contact history.

224 (59.7%) patients were male while 151 (40.3%) were female and the mean age of the patients was 58.83 years with a standard deviation of 16.46 years. Even though 76 demographic, laboratory, and clinical characteristics were available in the dataset, 14 biomarkers and two demographic variables were identified using feature ranking. Using two different feature ranking techniques, two different top-10 features were identified as most contributing features (Figure 2). Some features are found common to both the techniques resulting in 15 different features contributing most for early prediction of death.

The detailed description of 16 characteristics are listed in Table 1. It was found that gender, age, LDH, neutrophils (%), lymphocyte (%), hs-CRP, eosinophil (%), monocyte (%), INR, high sensitivity cardiac troponin I, NT-proBNP and albumin had statistically significant differences between the groups groups ($P < 0.05$), whereas serum sodium, serum chloride, APTT, MCHC variables were not significantly different ($P > 0.05$) among the two groups. Out of these 16 characteristics, 12 characteristics were observed statistically significant. Therefore, it was important to check the most useful variables for the early prediction of death.

### B. *UNIVARIATE LOGISTIC REGRESSION ANALYSIS OF VARIABLES SIGNIFICANTLY ASSOCIATED WITH DEATH*

To determine the independent variables associated with death, univariate logistic regression analysis was performed with Top-1, Top-2, and up to Top-10 features identified using two different techniques. It is clear from the Figure 3 that Top ranked 5 features produced highest AUC of 0.97 for data imputed using MICE algorithm while Top-ranked 3 features produced highest AUC of 0.95 for the data imputed using -1 (Figure 3). Table 2 shows the overall accuracies and weighted average performance for other matrices for different models using Top 1 to 10 features for 5-fold cross-validation using the logistic regression classifier along with the confusion matrices for each case.

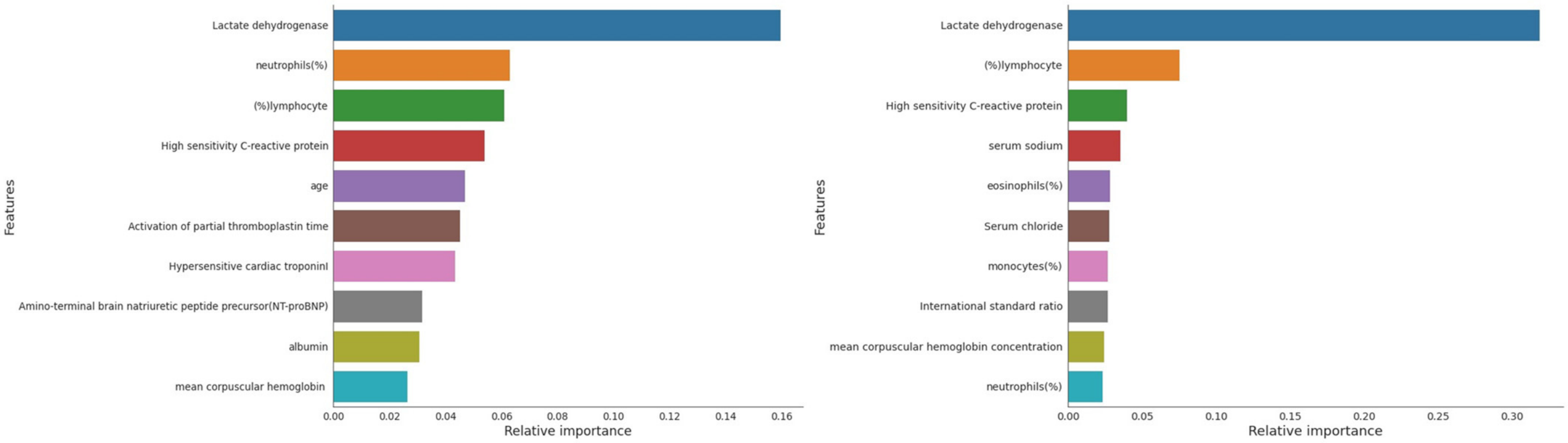

**Figure 2:** Comparison of the top-ranked 10 features identified using Multi-Tree XGBoost algorithm from data imputed using MICE (left) and (-1) (right).

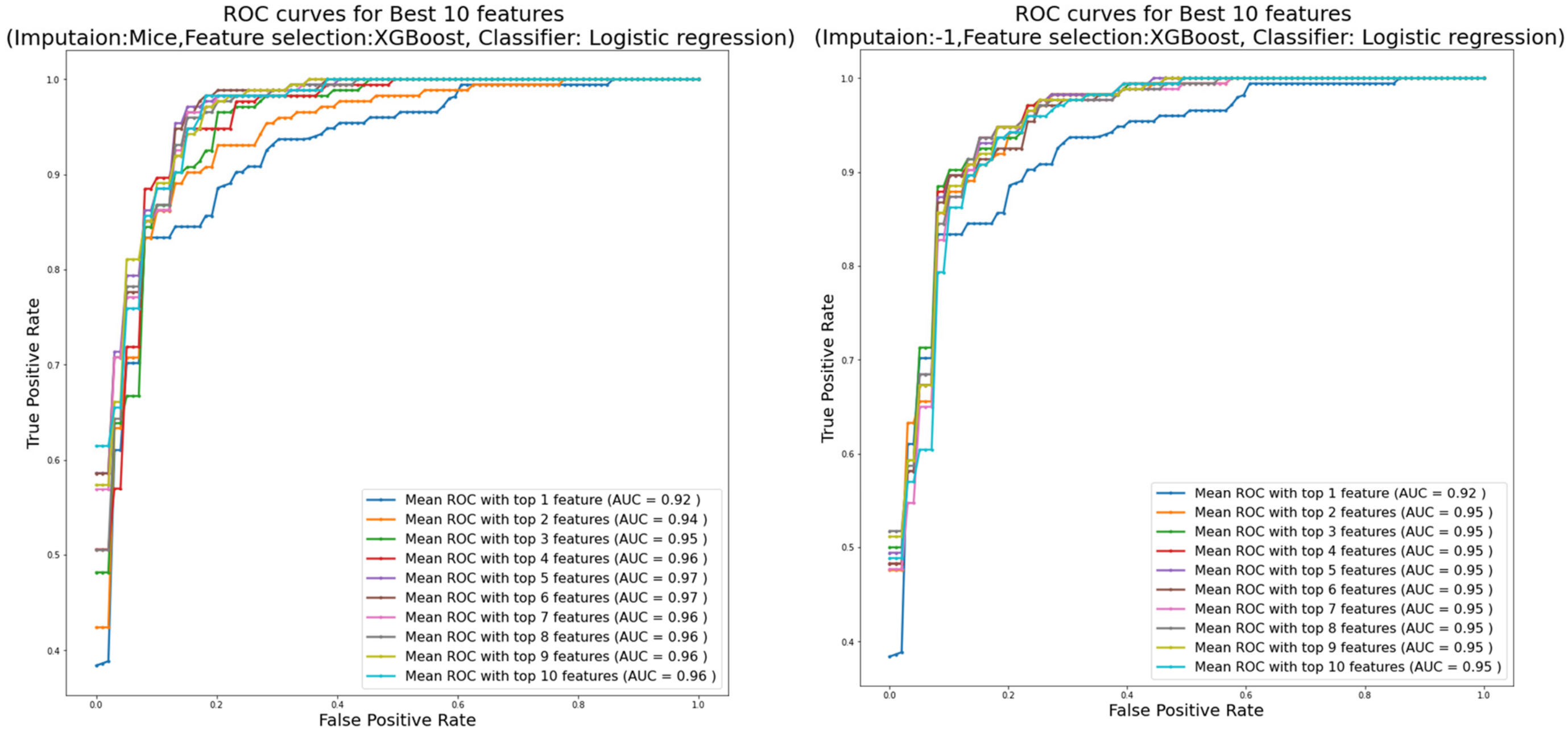

**Figure 3:** Comparison of the receive operating characteristic (ROC) plots for top-ranked 1 up to 10 features using the data imputation using MICE (left) and (-1) (right) while feature selection and classification techniques were same.

**Table 2:** Comparison of the average performance matrix and confusion matrix from five-fold cross-validation for top1 to 10 features using data imputation using (-1) (A) and mice (B).

| A | Weighted Average (95% confidence interval) | | | | Confusion Matrix | | | |
|---|---|---|---|---|---|---|---|---|
| | Sensitivity | Specificity | PLR | NLR | Death | | Survived | |
| | | | | | TP | FN | FP | TN |
| Top 1 feature | 87±3.92 | 87.4±3.01 | 7.4±4.1 | 0.15±0.1 | 142 | 32 | 14 | 187 |
| Top 2 features | 88.04±3.13 | 88±3.5 | 8.1±5.1 | 0.14±0.08 | 148 | 26 | 17 | 184 |
| Top 3 features | 90±3.8 | 88.9±3.78 | 9.3±6.9 | 0.12±0.09 | 155 | 19 | 19 | 182 |
| **Top 4 features** | **90.5±3.92** | **90.7±3.72** | **11.8±10.1** | **0.10±0.09** | **157** | **17** | **18** | **183** |
| **Top 5 features** | **90.1±3.6** | **90.03±3.5** | **10.5±7.9** | **0.11±0.086** | **155** | **19** | **18** | **183** |
| Top 6 features | 90.08±2.7 | 90±2.4 | 9.63±5.1 | 0.11±0.06 | 154 | 20 | 19 | 182 |
| Top 7 features | 89.8±2.3 | 90.16±3.4 | 10.5±7.5 | 0.12±0.05 | 156 | 18 | 21 | 180 |
| Top 8 features | 89.3±3.6 | 89.1±3 | 8.96±5.5 | 0.12±0.08 | 155 | 19 | 21 | 180 |
| Top 9 features | 89.6±3.2 | 88.9±3.5 | 9.06±6.2 | 0.11±0.07 | 153 | 21 | 20 | 181 |
| Top 10 features | 89.01±3.3 | 89.01±4 | 9.46±7.3 | 0.13±0.083 | 154 | 20 | 21 | 180 |

| B | Weighted Average (95% confidence interval) | | | | Confusion Matrix | | | |
|---|---|---|---|---|---|---|---|---|
| | Sensitivity | Specificity | PLR | NLR | Death | | Survived | |
| | | | | | TP | FN | FP | TN |
| Top 1 feature | 88.2±7.4 | 87.6±3.5 | 7.91±5.6 | 0.13±0.17 | 143 | 31 | 13 | 188 |
| Top 2 features | 87.7±4.4 | 87.01±3.5 | 7.37±4.6 | 0.14±0.11 | 145 | 29 | 17 | 184 |
| Top 3 features | 87.1±3.5 | 87±4.1 | 7.53±5.2 | 0.15±0.09 | 148 | 26 | 22 | 179 |
| Top 4 features | 89.2±2.8 | 89±3.2 | 8.93±5.6 | 0.12±0.07 | 155 | 19 | 22 | 179 |
| **Top 5 features** | **92±2.6** | **92±3** | **13.52±10.6** | **0.09±0.06** | **160** | **14** | **16** | **185** |
| **Top 6 features** | **92.3±2.45** | **92±4.1** | **15.86±16.5** | **0.085±0.06** | **162** | **12** | **17** | **184** |
| Top 7 features | 90.2±5 | 90.6±3.5 | 11.37±9.3 | 0.11±0.12 | 158 | 16 | 22 | 179 |
| Top 8 features | 89.9±4.8 | 90.2±3.8 | 11.02±9.3 | 0.11±0.11 | 158 | 16 | 23 | 178 |
| Top 9 features | 89.2±2.8 | 89.03±3.2 | 8.97±5.6 | 0.12±0.07 | 155 | 19 | 22 | 179 |
| Top 10 features | 88±3.4 | 89.6±3.7 | 9.82±7.5 | 0.14±0.08 | 156 | 18 | 23 | 178 |

Top-ranked 5 features using MICE data imputation showed better performance than the Top-ranked 4 features for the data imputed by (-1). Therefore, in the rest of the study, 5 Top-ranked MICE imputed independent variables: **L**actate dehydrogenase, **N**eutrophils (%), **L**ymphocyte (%), high sensitivity **C**-reactive protein and **A**ge (in short LNLCA) were used for nomogram creation and scoring technique development and validation.

### *C. DEVELOPMENT AND EVALUATION OF NOMOGRAM IN PREDICTING DEATH*

A multivariate logistic regression based nomogram for predicting early COVID-19 mortality was built using top-ranked five biomarkers that were found important both statistically and using ML based classifier (as shown in Table 1, 2 and Figure 3). The relationship between linear prediction of death and these biomarkers was evaluated using multivariable logistic regression which was reported in Table

3. Regression coefficient, z-value, standard error and its statistical significance along with 95% confidence interval were shown in Table 3. Z-value is the ratio of regression coefficient and its standard error. Typically z-value indicates the strong and weak contributors in logistic regression. The higher z-values (either positive/negative) represent a strong contributor while values close to zero represent a weak contributor. Therefore, out of 5 variables neutrophils (%) is not very strong predictor while age and Lactate dehydrogenase are strong contributor. A null hypothesis of particular regression coefficient can determine the p-value to relate the significance of a particular X-variable in relationship to the Y-variable. The X-variables for which p is less than 0.05, have significant relationship to Y-variables. This is also reflecting that the neutrophils (%) is weakly related to Y-variable. However, the logistic regression classifier shows that 5 variables outperforms than 4 variables. Therefore, no variable was discarded out of these 5-variables in developing the nomogram.

According to Figure 4, the calibration plot graphed closely toward the diagonal line both for internal and external validation which were indicative of the reliable model. It is evident from figure 5 that the net benefit of every single predictor model is positive until threshold of 0.85. This indicates that all of them contributed to the prediction of outcomes. Interestingly, the full model demonstrated the best performance which also confirmed the need to combine five predictors in the model.

As shown in Figure 6, the nomogram is comprised of 8 rows while row 1-5 are representing independent variables. For each variable, an assigned score was obtained by drawing a downward vertical line from the value on the variable axis to the "points" axis using COVID-19 patient data. The points of the five variables corresponds to score (row 6) and the scores were added up to the Total score, as shown in row 8. Then a line could be drawn from the "Total Score" axis to the "Prob" axis (row 7) to determine the death probability of COVID-19 patients. However, it is useful to derive the mathematical equations explaining the total score, linear prediction and death probability based on which the LNLCA score is calculated:

**Total points** = 0.0053375*lactate dehydrogenase (μ/L) - 0.02474*(Neutrophils (%)-98.7) - 0.12333*(Lymphocyte (%) - 60) + 0.0084375*hsCRP (mg/L) + 0.055844*age (years) (9)

**Linear prediction** $= -3.662636 + 0.0735038 \times \text{age (years)} + 0.0110451 \times \text{hsCRP} \left(\frac{\text{mg}}{\text{L}}\right) - 0.1624422 \times \text{lymphocyte(\%)} - 0.0327053 \times \text{neutrophils(\%)} + 0.0070514 \times \text{lactate dehydrogenase}\left(\frac{\text{u}}{\text{L}}\right)$ (10)

**Death probability**=1/ (1+exp (-Linear Prediction)) (11)

The corresponding probability of death for a given LNLCA score was determined from the model and is listed in Table 4. In particular, LNLCA score cut-off values of 10.4 and 12.65 were correspond to 5% and 50% of death probability, thus these values can be used to stratify COVID-19 patients into three groups: Low, moderate and high-risk groups. The death probability were less than 5%, between 5% and 50 % and more than 50 % for low risk group (LNLCA < 10.4), moderate risk group (10.4 ≤ LNLCA ≤ 12.65) and high risk group (LNLCA >12.65) respectively.

**Table 3:** The logistic regression analysis to construct the nomogram for death prediction.

| Outcome | Coef. | Std. Err. | z | P>\|z\| | [95% conf. | Interval] |
|---|---|---|---|---|---|---|
| Lactate dehydrogenase | .0070514 | .0017099 | 4.12 | 0.000 | .0037001 | .0104027 |
| Neutrophils | -.0327053 | .0568836 | -0.57 | 0.565 | -.1441951 | .0787845 |
| Lymphocyte | -.1624422 | .0806231 | -2.01 | 0.044 | -.3204607 | -.0044238 |
| High Sensitivity CRP | .0110451 | .0043462 | 2.54 | .011 | .0025267 | .0195635 |
| Age | .0735038 | .0185211 | 3.97 | 0.000 | .0372032 | .1098045 |
| _cons | -3.662636 | 5.65169 | -0.65 | 0.517 | -14.73975 | 7.414473 |

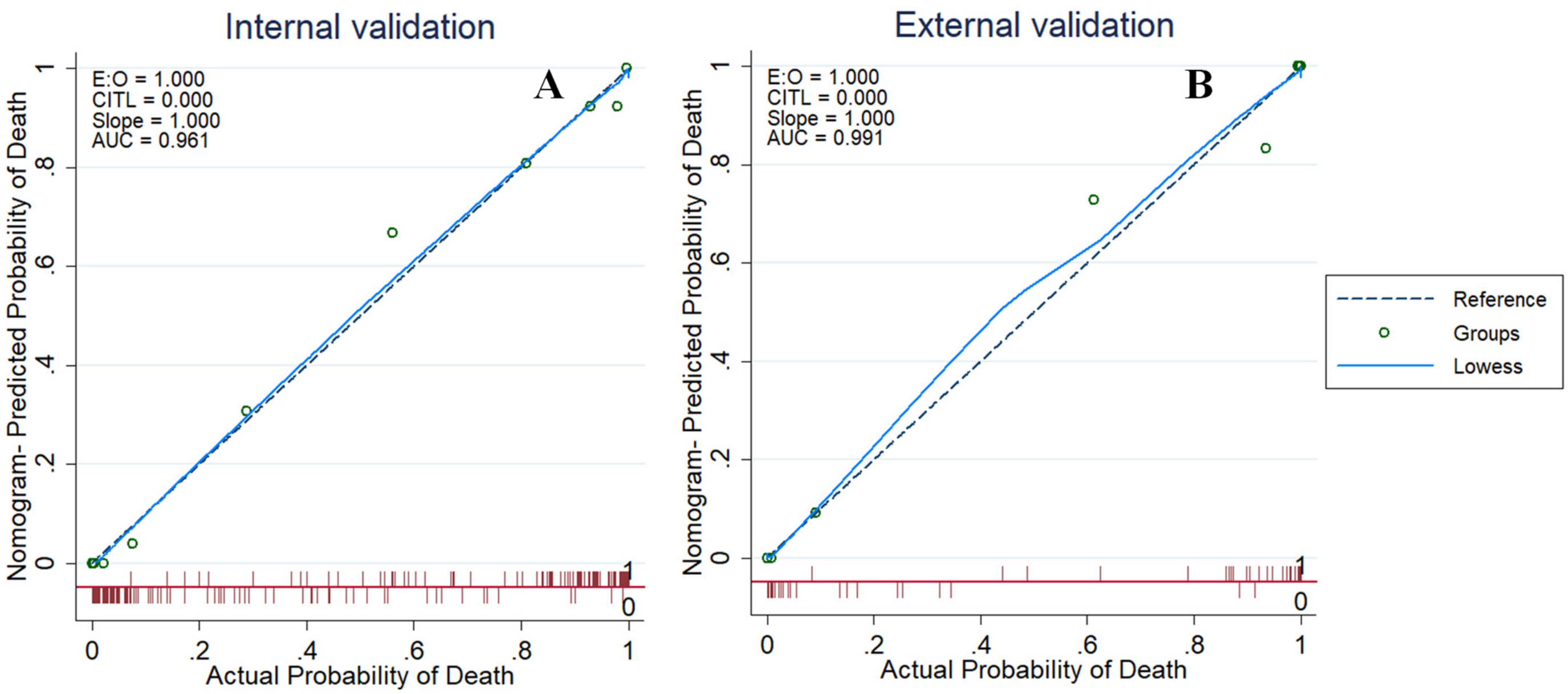

**Figure 4**: Calibration plot comparing predicted and actual death probability of patients with COVID-19: (A) represents the internal validation and (B) represents the external validation.

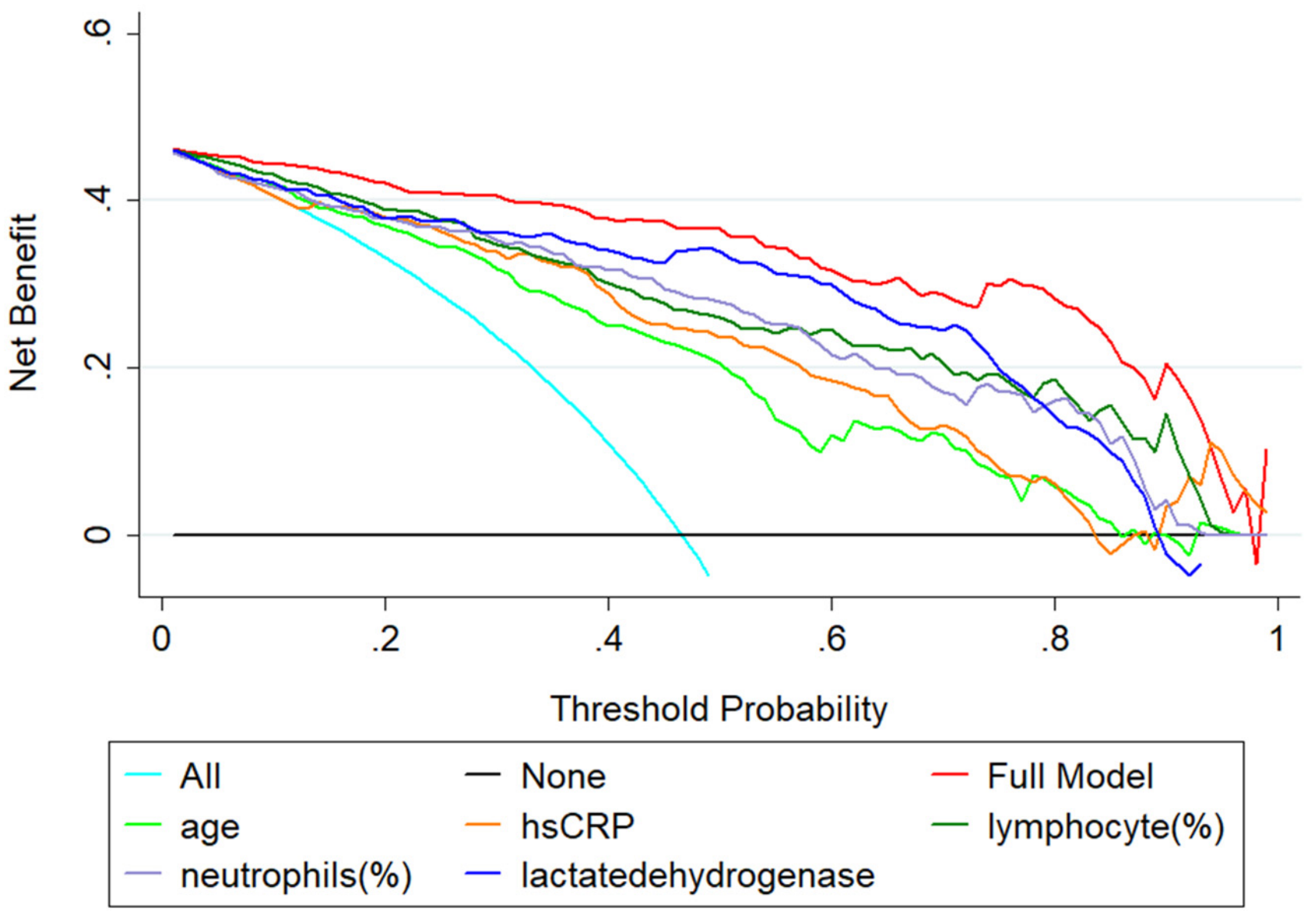

**Figure 5:** Decision curves analysis comparing different models to predict the death probability of patients with COVID-19. The net benefit balances the mortality risk and potential harm from unnecessary over-intervention for patients with COVID-19.

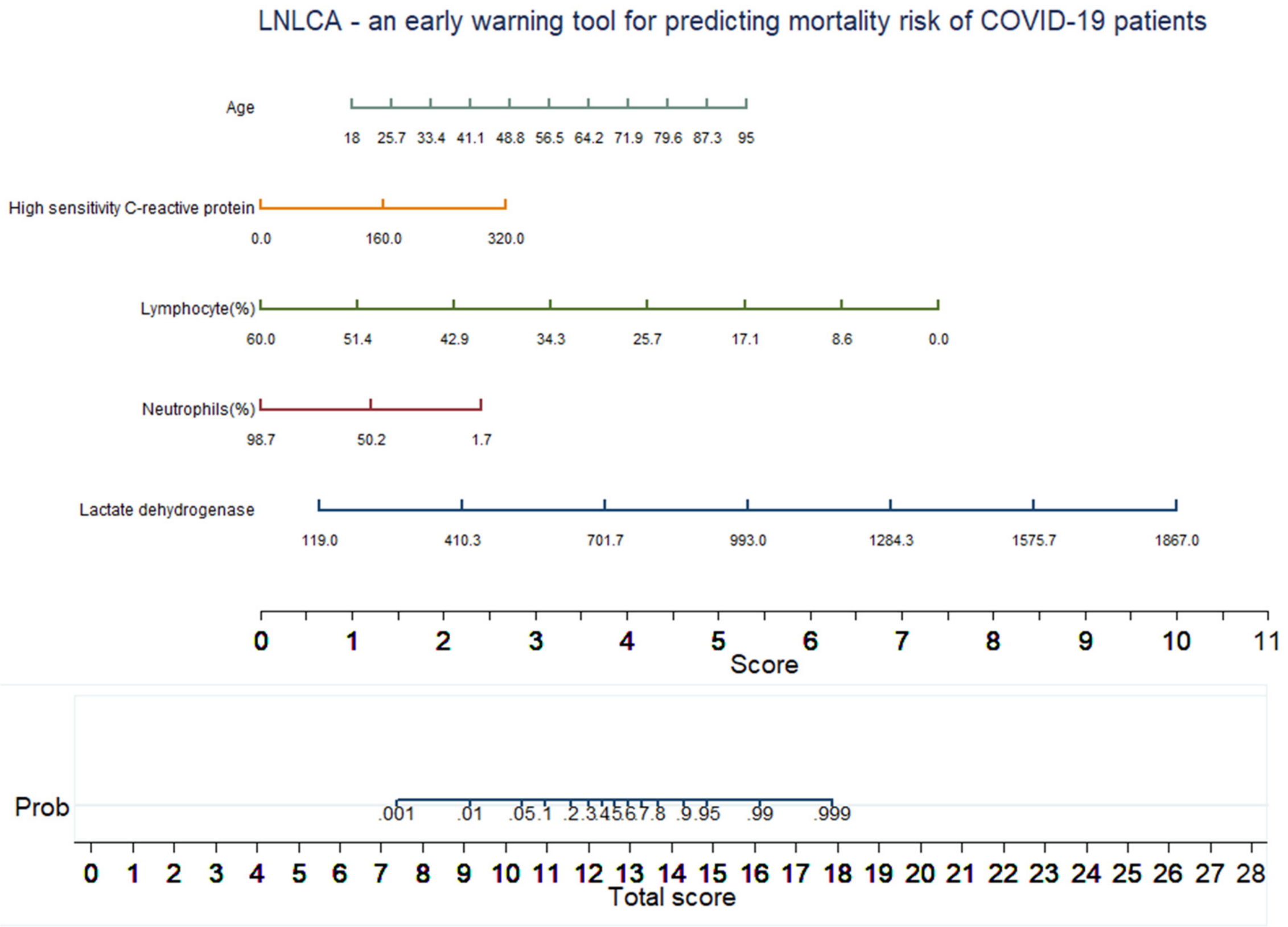

**Figure 6:** Multivariate logistic regression-based Nomogram to predict the probability of death. Nomogram for prediction of death was created using the following five predictors: Lactate Dehydrogenase, Neutrophils (%), Lymphocytes (%), High Sensitive C-reactive protein, and age.

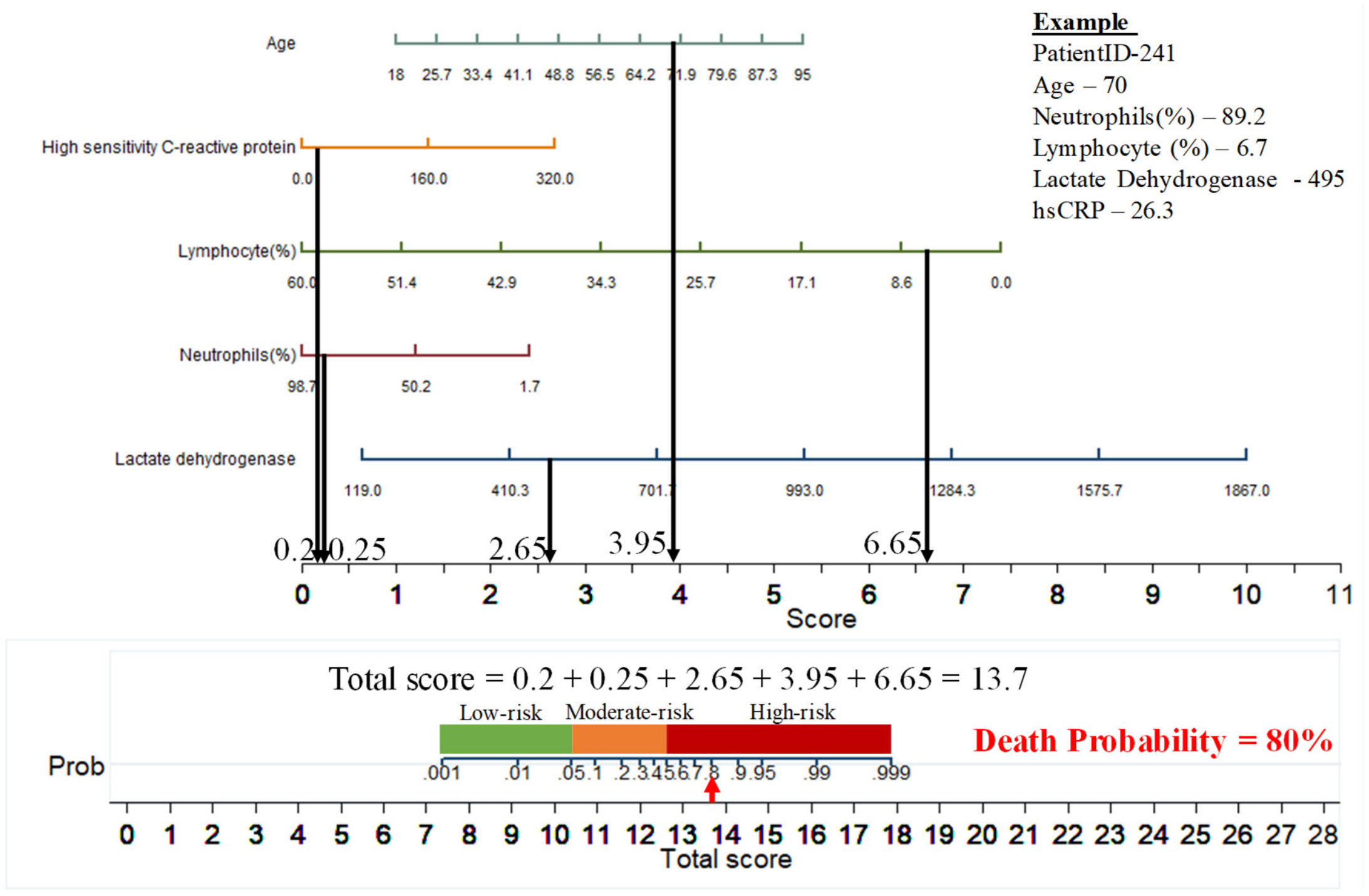

**Figure 7**: An example nomogram based score to predict the probability of death of a COVID-19 patient from test set (9-days before the actual outcome).

**Table 4:** LNLCA score from nomogram and corresponding death probability of COVID-19 patients

| Patient Group | LNLCA Score | Death probability |
|---|---|---|
| | 7.45 | 0.001 |
| **Low** | 9.2 | 0.01 |
| | **10.4** | **0.05** |
| | 10.95 | 0.1 |
| | 11.6 | 0.2 |
| **Moderate** | 11.99 | 0.3 |
| | 12.4 | 0.4 |
| | **12.65** | **0.5** |
| | 12.95 | 0.6 |
| | 13.3 | 0.7 |
| | 13.7 | 0.8 |
| **High** | 14.3 | 0.9 |
| | 14.8 | 0.95 |
| | 16.2 | 0.99 |
| | 17.85 | 0.999 |

### *D. PERFORMANCE EVALUATION OF THE MODEL*

Figure 7 shows an example nomogram based scoring system for a COVID-19 patient with the variable values at admission. Individual score for each predictors were calculated and added to produce total score and death probability was calculated to 80%. This can be done as early as 9 days before the death of the patient.

Furthermore, we have categorized the patients from training and testing subgroups into three subgroups (low, moderate and high-risk) by associating actual outcome with the predicted outcome using the LNLCA score. For training set (Table 5), the proportions of death were 0% (0/83) for low risk group, 22.6% (12/53) for moderate risk group and 88.1% (111/126) for high risk group while for test set (Table 6), the proportions of death were 0% (0/41) for low risk group, 22.7% (5/22) for moderate risk group and 94% (3/50) for high risk group. It was found that the true death rates were significantly different ($p<0.001$) among the three subgroups. Therefore, this nomogram based scoring technique can be used to early predict patients' outcome to categorize them into low, moderate and high-risk groups as shown in Table 4 and prioritize the moderate and high risk group patients.

There were 52 patients in the test set who had an outcome of death after different duration of hospital stay. Some patients were hospitalized in very late stages while some other patients were admitted in the early stages. The minimum, maximum,

**Table 5:** Association between different risk groups and actual outcome in the training cohort using Fisher exact probability test

| **Risk category** | **Outcome** | | **Overall** |
|---|---|---|---|
| | **Alive** | **Death** | |
| Low-risk | 83 (100.0%) | 0 (0%) | 83 (100.0%) |
| Moderate-risk | 41 (77.36%) | 12 (22.64%) | 53 (100.0%) |
| High-risk | 15(11.9%) | 111 (88.1%) | 126 (100.0%) |
| Overall | 139 (53%) | 123 (47%) | 262 (100.0%) |

P-value among three group is less than 0.001
P-value of Low-risk group vs Moderate-risk group is less than 0.001.
P-value of Low-risk group vs High-risk group is less than 0.001.
P-value of Moderate-risk group vs High-risk group is less than 0.001.

**Table 6:** Association between different risk groups and actual outcome in the Testing cohort using Fisher exact probability test

| **Risk category** | **Outcome** | | **Overall** |
|---|---|---|---|
| | **Alive** | **Death** | |
| Low-risk | 41 (100%) | 0 (0%) | 41 (100.0%) |
| Moderate-risk | 17 (77.27%) | 5 (22.73%) | 22 (100.0%) |
| High-risk | 3 (6%) | 47 (94%) | 50 (100.0%) |
| Overall | 61 (54%) | 52 (46%) | 113 (100.0%) |

P-value among three group is less than 0.001
P-value of Low-risk group vs Moderate-risk group is 0.0037.
P-value of Low-risk group vs High-risk group is less than 0.001.
P-value of Moderate-risk group vs High-risk group is less than 0.001.

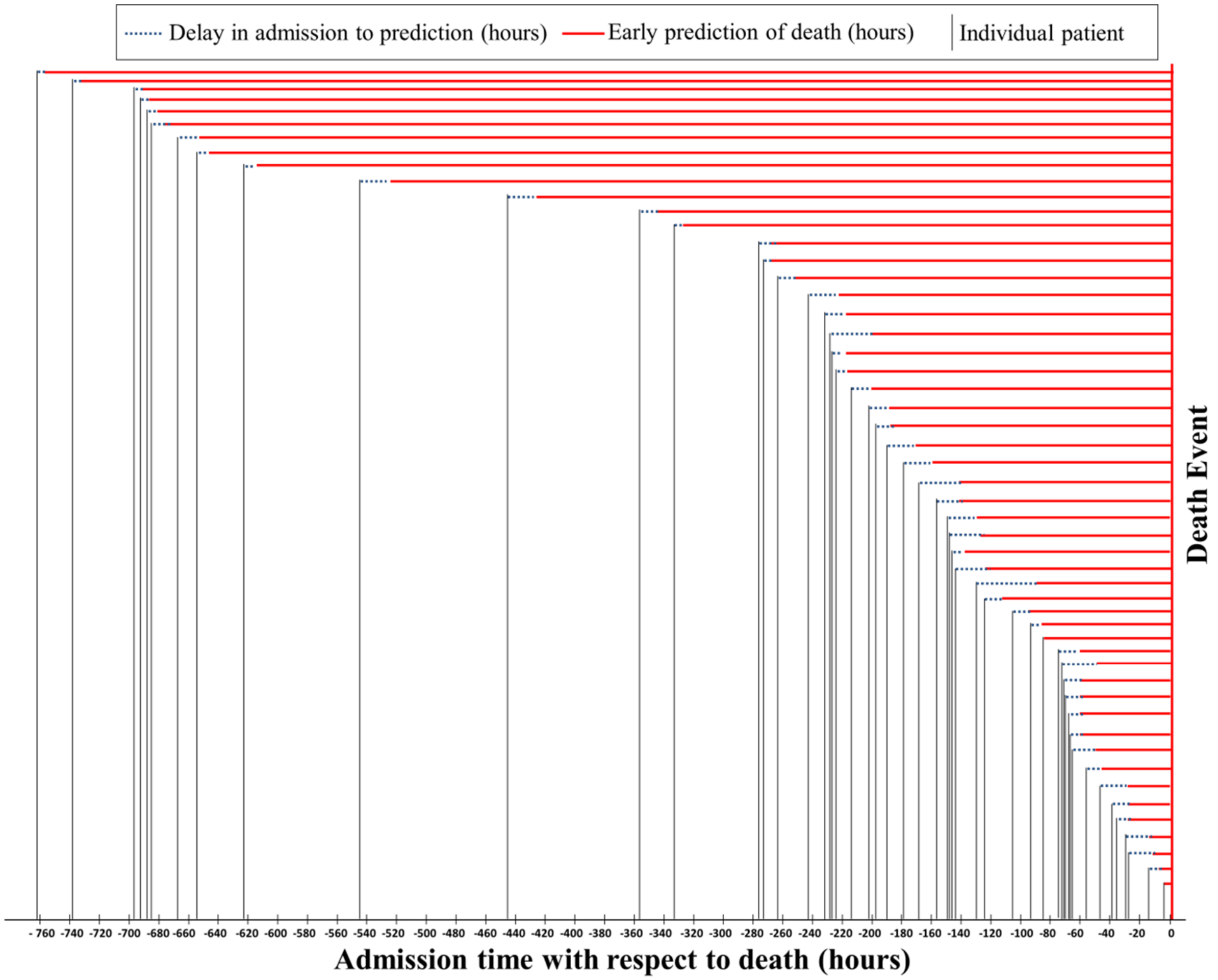

**Figure 8:** Estimation of the prediction of the patients' outcome for 52 test patients with death outcome. The model was trained on the data present at admission and multiple samples from a patient was used to predict the patient to be in high-risk group in the earliest time after admission. Note: '0' denotes the death outcome event for each patient and vertical lines represent the time of admission with respect to death. Solid red line start from the earlies prediction time point of death prediction and the dotted line represent the delay between admission and death prediction by the model using the LNLCA model.

mean (±standard deviation), and median of hospital admission to death for the test data set were 3.68, 760.92, 249.2±227.55 and 172.79 hours respectively. Most patients out of the 375 patients of the cohort had multiple blood samples taken throughout their hospital stay. LNLCA model based prediction score were calculated on the admission and also calculated for the next available samples and identified when the model is predicting the patient in high risk group in the earliest possible time after admission. Figure 8 shows the difference in hours between hospital admissions to the event of death and also shows when the model can predict the potential outcome with 100% accuracy. It was evident from Figure 8 that the model can predict the outcome of 52 patients within several hours after admission for most of the patients. The minimum, maximum, mean (±standard deviation), and median of model's high-risk prediction to death for the test data set were 3.68, 756.11, 239.85±228.56 and 156.36 hours respectively. The model can even predict 31.5 days in advance for a patient about the outcome with a probability of 97%. This early prediction suggests that, where a patient's condition deteriorates, the clinical route is able to give an early warning to clinicians several days in advance.

## DISCUSSION

Current study investigated the relationship between the disease severity and the clinical data. Ten predictors were identified by Multi-Tree XGBoost algorithm as death probability predictors based on the data acquired at hospital admission time. Two different prediction models were compared while missing data were imputed using -1 and using MICE algorithm. Ten different classification model trained, validated and tested for Top 1 to 10 features using two different techniques. It was observed from the AUC and performance matrices that the MICE based technique outperforms other approach with an AUC = 0.97 was achieved for 5 Top-ranked features. Then, a logistic regression based nomogram was developed using these five variables. An integrated score (LNLCA) with corresponding likelihood of death was obtained for the early stratification of COVID-19 patients based on the severity prediction. This can help to effectively the use the healthcare facilities without overloading their capability.

Age was identified as a key predictor of mortality in previous studies on Coronavirus family such as SARS [30], Middle East respiratory syndrome (MERS) [31] and COVID-19 [32]. This study has also concluded similar findings and this is because with the older age the immunosenescence and/or multiple medical conditions tend to make patients more prone to critical COVID-19 illness [19].

Yan et al. [16] showed that in patients with severe pulmonary interstitial disease, there is a significant increase of LDH and can be associated with indications for lung injury or idiopathic pulmonary fibrosis [33]. Consistent results from the previous research were also found in this study, in which critically ill patients with COVID-19 had elevated levels of LDH suggesting an increase in activity and severity of lung injury. LDH is an intracellular enzyme that leaks from damaged cells due to infection and viral replication leading to elevated levels in circulation.

Recently, Liu et al. [34] proposed that increased Neutrophil-to-Lymphocyte Ratio (NLR) can aid in the early prediction of the severity of COVID-19 illness. Both neutrophils and lymphocytes are critical components of the immune system and play very important role in host defense and clearing infections. Lymphopenia, medical condition due to lower number of lymphocytes in the blood, is a typical feature in COVID-19 patients, and may be a key factor in disease severity and mortality [35]. In this study, we have used neutrophils and lymphocytes percentage and similar to the previous studies have found that lower percentage of these two quantities were associated with severe COVID-19 patients. According to previous research, patients with community-acquired pneumonia have significant immune system activation and/or immune dysfunction leading to changes in these quantities [35]. In addition, on the event of immunosuppression and apoptosis of lymphocytes caused by specific anti-inflammatory cytokines, bone marrow circulates neutrophils [36], resulting in an increased NLR. However, in contrast to other models, it was observed in this study, both the parameters were small for high-risk patients.

Lu et al. [37] stated that CRP tested upon admission may assist in predicting confirmed or suspected short-term mortality associated with COVID-19. CRP is an acute phase protein formed by hepatocytes caused by leukocyte-derived cytokines induced by infection, inflammation or tissue damage [38-40]. Similar findings were found in this study where increased CRP rates were measured at admission for the high mortality risk COVID-19 patients. This indicated that these patients developed a serious lung inflammation or possibly a secondary bacterial infection, and clinical antibiotic treatment might be appropriate for those patients [21].

Non-survivors in our study had low lymphocyte and neutrophil percentages, higher age, hsCRP and LDH than those of survivors. In addition to the dysregulation of the coagulation system and/ immune system, it can be seen that COVID-19 severity was significantly linked to the inflammatory response to the infection. This could lead to other worse medical consequences like ARDS, septic shock and coagulopathy etc. Therefore, this kind of prognostic model will aid in the development of a rational and personalized therapeutic plan for the patients with critical illness.

Weng et al. [19] recently suggested that age, NLR, D-dimer and CRP were individual key predictors correlated with death probability. These key-predictors were used to create a nomogram for death prediction due to COVID-19. In our research, the five key predictors recorded at admission were chosen by the XGBoost feature selection to create a nomogram based prognostic model that exhibits excellent calibration and discrimination in predicting death probability of COVID-19 patients. It was also validated by an unseen validation cohort. Moreover, it was verified with multiple blood sample data collected from the patients during their hospital stay and the model holds valid for those cases as well. The AUC values for development and validation cohort showed a strong distinction of 0.961 and 0.991 respectively using the proposed nomogram, which is, to the best of our knowledge, outperforms any other nomogram based models for COVID-19 mortality prediction. In addition, this nomogram-derived LNLCA score offered a simple, easy-to-understand and interpretable early detection tool for stratifying the high-risk COVID-19 patients at admission and thereby assist their clinical management. COVID-19 patients were categorized into three risk groups with varying risk of death using LNLCA score measured and calculated at admission. Low-risk group cases could be isolated and treated in an isolation center while the moderate-risk patients could be treated isolation ward in a specialized hospital. On the other hand, patients in high-risk group could be under close monitoring and should be moved to critical medical services or ICU for urgent treatment if required.

This study has scope for further improvement, which will be carried out in the future work. Firstly, the study motivates the possibility of research on COVID-19 clinical data helping in early mortality prediction but the proposed machine learning method is purely data-driven and may vary if starting from different datasets. The model can be further improved with the help of a larger dataset. Secondly, the modelling principle adopted here is to have a minimal number of features for accurate predictions to avoid overfitting, which can be revised with several other models to identify any other sets of best features on a multi-center and multi-country data to produce a generalized model.

## CONCLUSION

In summary, based on multiple risk factors (Lactate Dehydrogenase, Neutrophils (%), Lymphocytes (%), High Sensitive C-reactive protein, and age), our developed nomogram can predict the prognosis of patients with COVID-19 with good discrimination and calibration. The model can predict the patient's outcome far ahead of the day of primary clinical outcome with very high accuracy. Therefore, the application of LNLCA would help clinicians make an efficient and optimized patient stratification management plan without overloading the healthcare resources and also reduce the death with improved and planned response. The authors also plan to further improve the performance of the model with the help of larger dataset with multi-center and multi-country data.

## COMPLIANCE WITH ETHICAL STANDARDS

**Disclosure of potential conflicts of interest**

The authors declare that they have no conflict of interest.

**Funding:** This publication was made possible by Qatar University Emergency Response Grant (QUERG-CENG-2020-1) from the Qatar University. The statements made herein are solely the responsibility of the authors.

**Ethical approval:** This article uses the clinical data which was made publicly available by Yan et al. [21]. Therefore, the authors of this study were not involved with human participants or animals. However, the original retrospective study carried out by Yan et al. [21] was approved by the Tongji Hospital Ethics Committee.